\documentstyle[12pt]{article}

\begin{document}

\newcommand{\eq}{\begin{equation}}
\newcommand{\en}{\end{equation}}
\newcommand{\eqa}{\begin{eqnarray}}
\newcommand{\ena}{\end{eqnarray}}

$\mbox{ }$
\vspace{-3cm}
\begin{flushright}
\begin{tabular}{l}
{\bf KEK-TH-361 }\\
{\bf KEK preprint 93 }\\
May 1993
\end{tabular}
\end{flushright}

\baselineskip18pt
\vspace{1cm}
\begin{center}
\Large
{\baselineskip26pt \bf 2D String Theory Coupled to \\
                       Quantum Gravity }
\end{center}
\vspace{1cm}
\begin{center}
\large
{\sc Nobuyuki Ishibashi}
\end{center}
\normalsize
\begin{center}
\begin{tabular}{l}
{\it Theory Group, KEK, Tsukuba, Ibaraki 305, Japan }
\end{tabular}
\end{center}
\vspace{2cm}
\begin{center}
\normalsize
ABSTRACT
\end{center}
{\rightskip=2pc 
\leftskip=2pc 
\normalsize
We consider self-avoiding Nambu-Goto open strings on a random surface. We have
shown that the partition function of such a string theory can be calculated
exactly. The string susceptibility for the disk is evaluated to be
$-\frac{1}{2}$.
We also consider modifications of the Nambu-Goto action which are exactly
soluble on a random surface.
\vglue 0.6cm}

\newpage
\section{Introduction}
\hspace{5mm}
String theory appears in many aspects of physics.
It is an old idea that QCD might be
represented as a string theory\cite{string}.
Also, three dimensional Ising model was argued
to be described as a three dimensional superstring theory\cite{3dis}.
Therefore, it is
important to search for consistent string theories in each dimension and
classify the universality classes of them. Such
string theories will be useful in describing various physical systems as
point particle field theories was.

However, constructing consistent string theories appropriate for such
applications is not an easy task.
It is always
possible to construct low dimensional string theories by compactifying
critical string theories, but such theories have features like the existence of
a massless spin two particle, which are not desirable in most cases. Hence we
should look for noncritical string theories. However, constructing a consistent
string theory preserving Lorentz or rotation invariance is not so easy.
The noncritical
Polyakov string was quantized by the authors of \cite{KPZ},
but their quantization is not
consistent in $d$ dimensional space-time with  $1 < d < 25$.
The string susceptibility becomes complex, if one naively applies the KPZ
formula to these cases.
Therefore one should overcome this ``$c=1$ barrier'' in order to
construct Polyakov string theories of physical interest.

There can be another approach to noncritical string theory, namely the direct
quantization
of the Nambu-Goto
action.
Usually the Nambu-Goto action is quantized by transforming it into the Polyakov
action. In 26 dimensional space-time, we can quantize the Nambu-Goto
string directly and it gives the same results as that of the quantization
\`{a} la Polyakov.
However, there is no reason to
believe in their equivalence in the noncritical cases. Therefore, it is
possible
that the Nambu-Goto string has no ``$c=1$ barrier'' when directly quantized.
The Nambu-Goto string is manifestly Lorentz or rotation invariant and it can be
useful in describing
various systems. Unfortunately the Nambu-Goto string has a highly nonlinear
action and intractable in general.
Here let us discuss the two dimensional self-avoiding Nambu-Goto string
as the simplest case in this direction.

In two dimensional string theories, the embedding of the worldsheet into
the space-time is
generically singular, involving folds. If one requires self-avoidingness, such
folds are forbidden and the Nambu-Goto action becomes trivial for closed
strings\cite{FKT}. However for open strings the theory is not so trivial.
For example, the partition
function corresponding to the disk graph of such an open string can be written
as
\eq
Z( \mu )= \sum_{\Gamma } \exp (-\mu A(\Gamma )).
\label{part}
\en
Here $\Gamma$ denotes the self-avoiding loop in the space-time corresponding
to the boundary of the open string graph and the sum is over such loops.
$A(\Gamma )$ is
the area of the space-time $\Gamma$ encloses. For the self-avoiding open
strings, the Nambu-Goto action depends only on the boundary of the graph, which
coincides with $A( \Gamma )$. This is a nontrivial sum and it is not sure if
this open string theory is consistent contrary to the usual noncritical
string case.

This self-avoiding open string theory may be applied to two
dimensional systems of interest, e.g. two dimensional QCD\cite{mig}, which
have been actively studied recently\cite{larN}.
The expectation value of the Wilson loop for two
dimensional pure Yang-Mills theory is known to be
\eq
<\mbox{Tr}_R P e^{\oint_C A_{\mu }dx^{\mu }}>= e^{-g^2C_2(R)A},
\en
where $g$ is the gauge coupling, $C_2(R)$ the quadratic Casimir operator for
representation $R$ of the gauge group, and $A$ is the area enclosed by the
loop. If one tries to incorporate dynamical quarks into the theory, one should
sum the above expectation value over the fluctuations of the loop $C$. Thus
we encounter a sum over loops similar to eq.(\ref{part}).
Eq.(\ref{part}) may correspond to two dimensional QCD with self-avoiding
quarks.

Anyway, it seems difficult to calculate the sum in eq.(\ref{part}).
In this paper, we will discuss this self-avoiding Nambu-Goto open string theory
on a random surface. In other words, we will consider the string theory coupled
to quantum gravity in the target space.
On a random surface,
it is possible to calculate the
partition functions of the open string theory exactly. We will show that the
string susceptibility is not complex contrary to the usual noncritical
string case. This fact implies the consistency of this string theory, at least
in the presence of quantum gravity in the target space.
In a sense, what we are
dealing with is related to two dimensional QCD with self-avoiding quarks
coupled to quantum gravity.

The organization of this paper is as follows. In section 2, we consider
self-avoiding random walks on a random surface as a warm-up. We will first
review
the techniques developed by Duplantier and Kostov\cite{DK} to calculate the
partition
function of self-avoiding random walks on a random surface. Using their
results,
we express the configuration sum of a loop on a random surface in terms of the
wave function of quantum gravity.
In section 3, we
go on to the case of self-avoiding strings and evaluate
the partition function. We will show that the string susceptibility for the
disk is
$-\frac{1}{2}$. We also discuss modifications of the Nambu-Goto
action. Section 4 contains concluding remarks.

\section{Self-avoiding random walks on a random surface}
\hspace{5mm}
The techniques needed for performing the sum in eq.(\ref{part}) over random
walks
on a random surface was developed by Duplantier and Kostov. They considered
self-avoiding walks on a random surface:
\eq
Z(\lambda ,m)=\sum_{\Gamma ,\mbox{metric}}e^{-ml-\lambda V},
\label{walk}
\en
where the sum is over self-avoiding walks $\Gamma$ and the space-time metrics
with the Boltzmann weight involving the total length $l$ of $\Gamma$ and the
volume $V$ of the space-time.
Let us first review their results\cite{DK}.

The ensemble of random surfaces was defined by the
continuum limit of the dynamical triangulation of the surfaces. If we restrict
the topology of the surface to that of the sphere, the partition function
$Z(\lambda ,m)$ in eq.(\ref{walk}) is discretized as
\eq
Z(\beta ,K)=\sum_{G}\frac{e^{-\beta |G|}}{S(G)}\sum_{\Gamma}K^{|\Gamma |}.
\label{disc}
\en
Here $G$ denotes a planar $\phi^3$ graph, $|G|$ is the number of vertices of
the graph $G$ and $S(G)$ is the symmetry factor.
The dual of $G$ describes a triangulated surface. $\Gamma$ is a self-avoiding
random walk on the graph $G$. $\beta $ and $K$ correspond to the parameters
$\lambda $ and $m$ respectively, in the continuum limit.

Duplantier and Kostov
rewrote eq.(\ref{disc}) in terms of the partition function $G_n$
of random graphs with $n$ external legs
\eq
G_n(\beta )=\sum_{n \mbox{~leg~planar~} G}e^{-\beta |G|}.
\en
Here, let us concentrate on the case where $\Gamma$ in eq.(\ref{disc}) is a
loop. Then $\Gamma$ divides the surface into two parts. $Z(\beta ,K)$ can be
expressed by two $G_n$'s representing these two parts as
\eq
Z(\beta ,K)=\sum_{m,n}\frac{1}{m+n}\frac{(m+n)!}{m!~n!}(e^{-\beta }K)^{m+n}
                       G_m(\beta )G_n(\beta ).
\label{free}
\en
$G_n(\beta )$ was calculated using the large-$N$ limit of the matrix model and
it has an integral expression \cite{BIPZ}:
\eq
G_n(\beta )=\int^{2b}_{2a}d\lambda \rho (\lambda )\lambda ^n.
\en
Here $\rho (\lambda )$ is the density of eigenvalues of the $N\times N$ matrix
given in \cite{BIPZ}. Substituting this expression, we obtain
\eq
Z(\beta ,K)=-\int^{2b}_{2a}d\lambda _1 \int^{2b}_{2a}d\lambda _2
            \rho (\lambda _1)\rho (\lambda _2)
            \ln (1-e^{-\beta }K(\lambda _1+\lambda _2)).
\label{int}
\en
The singular behaviour of $Z(\beta ,K)$ can be seen from this integral
representation. There is a critical point $\beta _c$ for $\beta $ which
corresponds to the pure gravity critical point. The critical point for $K$ is
at
\eq
K_c=e^{\beta _c}/4b_c,
\en
where $b_c$ is the value of $b$ at $\beta =\beta_c$.
When $K$ approaches $K_c$ simultaneously as $\beta $ approaches $\beta_c$,
\footnote{Here we consider only the {\it dilute phase} in \cite{DK}.}
the integral in eq.(\ref{int}) diverges because the point
$\lambda _1=\lambda _2=2b$ in the integration region approaches the singularity
of the logarithm in the integrand. Loops with infinite length dominate at this
critical point.

Eq.(\ref{int}) is the main result of Duplantier and Kostov, which we will
use in this paper.
In the rest of this section, we will calculate $Z(\lambda ,m)$ in
eq.(\ref{walk}) using eq.(\ref{int}).
In order to extract the singular part and take the continuum limit,
it is convenient to rewrite
eq.(\ref{int}) as
\eq
Z(\beta ,K)=\int^{2b}_{2a}d\lambda _1 \int^{2b}_{2a}d\lambda _2
            \rho (\lambda _1)\rho (\lambda _2)
            \lim_{\epsilon \rightarrow 0}\partial _{\epsilon }
            (\frac{1}{\Gamma (\epsilon )}
            \int^{\infty}_{0}dtt^{\epsilon -1}
            e^{-t(1-e^{-\beta }K(\lambda _1+\lambda _2))}),
\en
and change the variables as
\eq
e^{-\beta }=e^{-\beta _c}(1-\lambda \delta ^2),\;
K=K_c(1-m\delta ),\;
t=l/\delta.
\en
In the continuum limit, $\delta $, which is supposed to be the lattice spacing,
approaches zero. We use $Z(\beta ,K)$  in this limit to define the sum in
eq.(\ref{walk}) with $m$ and $\lambda $ as above. As $\delta \rightarrow 0$,
\eq
Z(\beta, K)\rightarrow \lim_{\epsilon \rightarrow 0}\partial _{\epsilon }
          (\frac{\delta ^{5-\epsilon }}{\Gamma (\epsilon )}
            \int^{\infty}_{0}dll^{\epsilon -1}\Psi (l)\Psi (l)e^{-ml}),
\label{limit}
\en
where
\eq
\Psi (l)=\lim_{\delta \rightarrow 0}\delta ^{-5/2}
         \int_{2a}^{2b}d\mu \rho (\mu )
         e^{-\frac{l}{\delta }\frac{2b_c-\mu}{4b_c}}.
\en

The right hand side of eq.(\ref{limit}) has a natural interpretation as
follows.
$\Psi (l)$ is the partition function of two dimensional gravity for the disk
with the boundary length $l$, which can be regarded as the wave function of
quantum gravity\cite{MSS}.
The right hand side of eq.(\ref{limit}) can be considered as a regularized form
of
\eq
\int^{\infty}_{0}\frac{dl}{l}\Psi (l)\Psi (l)e^{-ml}.
\en
When $m=0$, this formally coincides with the inner product of
the wave functions.
The above result shows that such an inner product gives the number of
configurations of self-avoiding loops on a random surface. Since the
wave function in quantum gravity does not depend on time, but rather is
something integrated over the time variable, it is easy to understand
intuitively that the inner product would give the number of ways of
taking a time slice on a random surface. The time slice we are considering
here is a self-avoiding random walk in the space-time.

Using $\rho (\mu )$ in \cite{BIPZ}, we can
evaluate $\Psi (l)$ up to an overall constant factor as,
\eq
\Psi (l)=l^{-5/2}(1+\sqrt{\lambda }l)e^{-\sqrt{\lambda }l},
\label{psi}
\en
after a rescaling of the variable $\lambda $. Substituting this expression,
it is easy to calculate the right hand side of eq.(\ref{limit}).
$\Psi (l)\Psi (l)e^{-ml}$ in the integrand can be expressed as a finite sum
of terms of the form $l^{n}e^{-(m+2\sqrt{\lambda})l}$ with $n$ an integer.
The following
formula is useful in evaluating the right hand side of eq.(\ref{limit}):
\eqa
\lim_{\epsilon \rightarrow 0}\partial _{\epsilon }
          (\frac{\delta ^{-\epsilon }}{\Gamma (\epsilon )}
            \int^{\infty}_{0}dll^{\epsilon -1}
           l^{n}e^{-(m+2\sqrt{\lambda})l})
&=&
-(l^{n}e^{-(m+2\sqrt{\lambda})l})_0\ln (m+2\sqrt{\lambda})
\nonumber
\\
& &
\hspace{7mm}
+(\mbox{terms~analytic~in~}m).
\label{for}
\ena
Here $(f(l))_0$ denotes the coefficient $a_0$ of $l^0$ in the Laurent expansion
$f(l)=\sum a_nl^n$.\footnote{$(f(l))_0$ appeared in the definition of an
inner product of the wave functions in \cite{MSS}.}
Thus we obtain as $Z(\lambda ,m)$,
\eq
-(\Psi (l)\Psi (l)e^{-ml})_0\ln (m+2\sqrt{\lambda})+
(\mbox{terms~analytic~in~}m).
\label{ana}
\en

The result
includes the nonuniversal part which is analytic in $m$ and vanishes when
differentiated sufficiently many times. It is the
contribution of very short loops. Therefore, the first term in eq.(\ref{ana})
is the universal part which should be taken as the continuum limit. We can
show that it is really universal and does not change if one changes
the way of discretization by making use of quadrangles etc. instead of
triangles.
Discarding the nonuniversal part, we obtain
as the continuum limit:
\eq
Z(\lambda ,m)=\mbox{[} \frac{1}{5!}(m+2\sqrt{\lambda })^5
                      -\frac{2}{4!}\sqrt{\lambda }(m+2\sqrt{\lambda })^4
                      +\frac{\lambda }{3!}(m+2\sqrt{\lambda })^3
              \mbox{]}
              \ln (m+2\sqrt{\lambda }).
\en

\section{Self-avoiding open strings on a random surface}
\hspace{5mm}
It is straightforward to apply the techniques in the previous section to
self-avoiding open strings on a random surface. The configuration
sum we have to deal with is as follows:
\eq
Z(\lambda ,\mu )=\sum_{\Gamma ,\mbox{metric}}e^{-\mu A-\lambda V}.
\label{stra}
\en
In this case, the Boltzmann weight involves the area $A$ enclosed by the
self-avoiding walk $\Gamma $. Let us restrict ourselves to the case where the
topology of the space-time is that of the sphere.
If one discretizes this sum as in the previous
section, it can again be rewritten by two $G_n$'s each of which corresponds to
the outside and the inside of the loop $\Gamma $.\footnote{We consider that
$\Gamma $ is oriented and we can define the inside and the outside. }
However, this time the cosmological constant in the $G_n$ corresponding to the
inside is shifted because of the term $\mu A$ in the Boltzmann weight.

Therefore, after the same procedure as in section 2, we can derive
\eq
Z(\lambda ,\mu )=\lim_{\epsilon \rightarrow 0}\partial _{\epsilon }
          (\frac{\delta ^{-\epsilon }}{\Gamma (\epsilon )}
            \int^{\infty}_{0}dll^{\epsilon -1}
            \Psi _{\mu +\lambda }(l)\Psi _{\lambda }(l)).
\label{inn}
\en
Here $\Psi _{\lambda }(l)$ denotes the disk partition function of two
dimensional quantum gravity with the cosmological constant $\lambda $. The
right
hand side of eq.(\ref{inn}) can be considered as
the regularized inner product of two wave functions with different
cosmological constants. The calculation of eq.(\ref{inn}) goes exactly as
in section 2.
Substituting eq.(\ref{psi}), one can again use the formula eq.(\ref{for}) with
a
slight modification. In order to select the universal part, we should insert
$e^{-ml}$ into the integrand of eq.(\ref{inn}), discard the terms analytic in
$m$ and take $m$ to be $0$ in the end.
Eventually we obtain
\eq
Z(\lambda ,\mu )=\mbox{[}-\frac{1}{30}(\sqrt{\lambda }+\sqrt{\lambda +\mu })^5
                 +\frac{1}{6}\sqrt{\lambda }\sqrt{\lambda +\mu }
                  (\sqrt{\lambda }+\sqrt{\lambda +\mu })^3\mbox{]}
                 \ln (\sqrt{\lambda }+\sqrt{\lambda +\mu }).
\label{stpt}
\en

When $\mu \gg \lambda $, the typical area of the worldsheet of the open string
is much smaller than the typical area of the space-time. Then the compactness
of the space-time becomes irrelevant to the fluctuations of the open string and
we expect that the partition function scales as
\eq
Z(\lambda ,\mu )\sim \mu ^{-\Gamma _{disk}+2}.
\en
This $\Gamma _{str.}$ should be taken as the definition of the string
susceptibility for the disk. When $\mu \gg \lambda $, eq.(\ref{stpt}) gives
\eq
Z(\lambda ,\mu )\sim \mu ^{\frac{5}{2}}\ln (\mu ),
\en
and $\Gamma _{disk}=-\frac{1}{2}$. Therefore the string susceptibility of this
two dimensional string theory is not complex and we do not encounter the
inconsistency contrary to the case of the usual noncritical string.

It is possible to generalize our self-avoiding Nambu-Goto string on a random
surface as follows.
The action of the Nambu-Goto string is the area of the worldsheet.
However any reparametrization invariant quantity is conceivable as a string
action.
Let us perturb the Nambu-Goto action by reparametrization
invariant operators $O_i$.
Then the sum in eq.(\ref{part}) will be modified to be
\eq
Z(\lambda ,\mu ,g_i)=\sum_{\Gamma }
                 e^{-\mu A-\sum g_iO_i}.
\en
and we obtain instead of eq.(\ref{inn}),
\eq
Z(\lambda ,\mu ,g_i)=\lim_{\epsilon \rightarrow 0}\partial _{\epsilon }
          (\frac{\delta ^{-\epsilon }}{\Gamma (\epsilon )}
            \int^{\infty}_{0}dll^{\epsilon -1}
            \Psi _{\mu +\lambda ,g_i}(l)\Psi _{\lambda }(l)).
\label{inn2}
\en
Here $\Psi _{\mu +\lambda ,g_i}(l)$ is the wave function of quantum gravity
with the action perturbed by $\sum g_iO_i$. We will show that such
perturbations to the Nambu-Goto action can change the critical behaviours of
the
theory.

By choosing $O_i$ to be the reparametrization invariant observables of
2d quantum gravity in \cite{DS}, and fine tuning $g_i$,
$\Psi _{\mu +\lambda ,g_i}(l)$ becomes the wave function of quantum gravity in
the multicritical phase\cite{KAZ}.
The explicit form of such a wave function in the
$m$-th multicritical point is\footnote{Here we take the conformal background in
\cite{MSS}, because we want $\mu$ to couple to the area of the worldsheet.}
\eq
\Psi _{\mu +\lambda ,g_i}(l)=
l^{-1}(\sqrt{\mu +\lambda })^{m-\frac{1}{2}}
K_{m-\frac{1}{2}}(\sqrt{\mu +\lambda }l),
\label{mfun}
\en
where $K_{m-\frac{1}{2}}$ is the Bessel function\cite{MSS}.
$m=2$ corresponds to the
pure gravity and the critical points with $m>2$ can be reached starting
from the pure
gravity. If one perturbs the Nambu-Goto action so that the wave function
$\Psi _{\mu +\lambda ,g_i}(l)$ which appears in eq.(\ref{inn2}) is the
multicritical wave function eq.(\ref{mfun}), one can obtain a new class of
string theories. We may call such a string the ``multicritical string''.
The string susceptibility for the disk of the multicritical string theory can
be
calculated using eq.(\ref{inn2}) and eq.(\ref{mfun}) as
\eq
\Gamma_{disk}
=
\left\{ \begin{array}{ll}
\frac{-m+1}{2} &(\mbox{if $m=2,4$ or odd})
\nonumber
\\
\frac{-m+4}{2} &(\mbox{otherwise}).
\end{array}
\right.
\en

\section{Conclusions}
\hspace{5mm}
We have shown that the two dimensional self-avoiding Nambu-Goto string is
exactly solvable when it is coupled to quantum gravity. The string
susceptibility
can be calculated and we obtain a real value.
Usually we quantize string theory \`{a} la Polyakov. If one tries to quantize
the two dimensional Polyakov string, one either spoils Lorentz or rotation
invariance, or encounters a complex string susceptibility.
The string theory considered here is different from the usual noncritical
string
theory in the following three points. Firstly here we deal with the direct
quantization of the Nambu-Goto action. Second, the string here is
self-avoiding.
And the last point is that the string is coupled to quantum gravity. Any of
these points can be the reason why the string theory has a real string
susceptibility contrary to the usual noncritical string.

We have also constructed new string models on a random surface by perturbing
the
Nambu-Goto action. We have shown that multicritical phases can be reached by
such perturbations. It is possible to calculate multiloop amplitudes
and obtain the mass spectrum of such string theories. We will report on this
problem elsewhere.

\section*{Acknowledgements}
We would like to thank K.Higashijima, Y.Okada, N.Tsuda, Y.Yamada, T.Yukawa
and other
members of KEK theory group for useful discussions and encouragements.

\end{document}